\documentclass[prl,a4paper,twocolumn,amsmath,]{revtex4-1}
\usepackage{graphicx}
\usepackage{epstopdf}
\usepackage{amsmath}
\usepackage{multirow}
\usepackage{color}

\begin{document}

\title{Fragmentation of molecules by virtual photons from remote neighbors}

\author{Lorenz S. Cederbaum}
\email[]{E-mail:  Lorenz.Cederbaum@pci.uni-heidelberg.de}
\affiliation{Theoretische Chemie, Physikalisch-Chemisches Institut, Universit\"at Heidelberg, Im Neuenheimer Feld 229, Heidelberg D-69120, Germany}
\date{\today}

\begin{abstract}

It is shown that a molecule can dissociate by the energy transferred from a remote neighbor. This neighbor can be an excited neutral or ionic atom or molecule. If it is an atom, the transferred energy is, of course, electronic and in the case of molecules it can also be vibrational. Explicit examples are given which demonstrate that the transfer can be highly efficient at distances where there is no bonding between the transmitter and the dissociating molecule.  
\end{abstract}

\maketitle

The fragmentation of a molecule through the absorption of light is called photodissociation. Being a fundamental process in nature, photodissociation attracted a vast amount of research, see, for instance, \cite{Schinke_Book,Rau_Book,Helen_Book,Salman_Book,Bio_Book,PD_Gas,PD_AA} and references therein. As bonds are broken in this process, its importance for chemistry, molecular science and astrophysics cannot be denied. As a consequence, nowadays, photodissociation is found to be important for modeling the chemistry of nearly every type of astrophysical region, see \cite{PD_AA} and references therein.

In general, the molecule undergoing photodissociation is not isolated, and we may ask what is the impact of a neighbor on this process. We concentrate here on the case where the molecule and its neighbor are well separated and do not posses a chemical bond such that if the photon is absorbed by the molecule, the photodissociation process is only little affected. We shall show, however, that if the impinging photon is absorbed by the neighbor, the molecule can still undergo fragmentation. 

Let us consider the following scenario from the point of view of the neighbor. The neighbor is excited or ionized either by an impinging photon or by the impact of another particle, like an electron or ion, and now possesses excess energy. If this excess energy is smaller than the ionization potential of the molecule, but larger than its dissociation energy, which is typically a rather large range of energy \cite{Schinke_Book,Rau_Book,Helen_Book,PD_AA}, then the neighbor can relax and transfer its excess energy to the molecule which will dissociate. 

The rate of this relaxation process is determined by the golden rule 

\begin{equation}\label{Golden_Rule}
\Gamma = 2\pi \sum_{f} \lvert \langle {\Psi_i}| V | {\Psi_f}\rangle   \rvert^2,
\end{equation}
where $V$ is the interaction between the molecule and its neighbor. The wavefunctions ${\Psi_i}$ and ${\Psi_f}$ describe as usual the initial and final states of the process in the absence of this interaction. The initial state is given by the product ${\Psi_i} = \phi_i^{N} \phi_i^{M}$ and the final state by ${\Psi_f} = \phi_f^{N} \phi_f^{M}$, where $N$ stands for the neighbor and $M$ for the molecule. Initially, the molecule is in its electronic ground state and vibrationally in any state of interest (usually the ground state) $\phi_i^{M}$ and the neighbor is an excited or ionized state $\phi_i^{N}$ as discussed above. After the process, the neighbor is in an energetically lower state, usually its ground state (neutral or ionic) $\phi_f^{N}$ and the molecule in the energy normalized continuum state $\phi_f^{M}$ describing the fragmented molecule. The sum over the final states also includes possible different states of the fragments. 

To allow not only for electronic, but also for vibrational to vibrational and vibrational to electronic and \textit{vice versa} energy transfer, the interaction $V$ contains the Coulomb interaction among all charged particles, electrons and nuclei. Let the electronic and nuclear coordinates of the neighbor relative to its center of mass be ${\bf{r}}_i$ and ${\bf{R}}_k$ and those of the molecule relative to its center of mass be ${\bf{r'}}_j$ and ${\bf{R'}}_l$. Expanding the interaction $V$ in inverse powers of the distance between the two centers of mass $R$ provides the leading contributing term \cite{Vibr_ICD} :

\begin{align}\label{Expansion}
 \frac{-3 ({\bf {u}} \cdot {\bf {\hat{D}}}^N)({\bf {u}} \cdot {\bf {\hat{D}}}^M) +  {\bf {\hat{D}}}^N \cdot {\bf {\hat{D}}}^M }{R^3} + O(\frac{1}{R^4}),
\end{align}
where {\bf{u}} is the unit vector connecting the two centers of mass, and 

\begin{align}\label{Dipole_Operators}
\nonumber &{\bf {\hat{D}}}^N = -\sum_{i}{\bf {r}}_i + \sum_{k}Z_k{\bf{R}}_k \\& {\bf {\hat{D}}}^M = -\sum_{j}{\bf r}'_j + \sum_{l}Z'_l{\bf{R}}'_l 
\end{align}
are the dipole operators of the neighbor and the molecule including all charged particles. The Z indicate nuclear charges.

\section{Results}

We now return to the golden rule (\ref{Golden_Rule}). It is straightforward to express its matrix element needed for the golden rule. Averaging over the orientations of the molecule and its neighbor leads to $\Gamma = \frac{4\pi}{3R^6}  \sum_{f'} \lvert {\bf D}_{i,f}^N\rvert^2 \lvert{\bf D}_{i,f'}^M \rvert^2$, where ${\bf {{D}}}_{i,f}^N = \langle \phi_i^{N}|{\bf {\hat{D}}}^N |\phi_f^{N}\rangle$ and similarly for the molecule. 

The dipole matrix elements entering the expression for the rate are closely related to measurable quantities and can be conveniently replaced by them. The Einstein coefficient, i.e., the inverse of the radiative lifetime, $A_{i,f}^N$ of the energy $E_{i,f}$ releasing transition of the neighbor reads \cite{Thorne_Book}
\begin{align}\label{Einstein_Coefficient}
A_{i,f}^N = \frac{4E_{i,f}^3}{3\hbar^4 c^3} \lvert{\bf D}_{i,f}^N\rvert^2,  
\end{align}
where c is the speed of light, and of central importance to this work, the molecular dipole matrix element determines the \textit{photodissociation cross section} of the molecule \cite{Rau_Book}:
\begin{align}\label{PD_CS}
\sigma_{_{PD}}^M(E_{i,f}) = \frac{4\pi^2}{3\hbar} \frac{E_{i,f}}{c}  \sum_{f'} \lvert {\bf D}_{i,f'}^M \rvert^2. 
\end{align}

We can view the process discussed above as follows. The neighbor possessing excess energy can relax by emitting a virtual photon which dissociates the molecules. We remind that the excess energy itself can be deposited by a photon or by the impact with another particle. At large distances $R$, the relaxation rate takes on the appearance 
\begin{align}\label{Gamma}
\Gamma_{_{VPD}} = \frac{3\hbar^5}{4\pi} \left(\frac{c}{E}\right)^4 \frac{A^{N}\sigma_{_{PD}}^M}{R^6}, 
\end{align}
where indices have been removed for simplicity and the subscript {\tiny $VPD$} has been added for later purpose. The rate is faster the faster is the radiative decay of the neighbor and the larger is the molecule's photodissociation cross section. The excess energy and the distance to the molecule influence the rate sensitively. The lifetime of the initial state of the neighbor due to the described relaxation process is $\tau_{_{VPD}} = \hbar / \Gamma_{_{VPD}}$. Of course, one is interested in cases where this lifetime is faster than that without the presence of the molecule. We shall see that this applies in many situations. A schematic picture of the process is shown in Figure 1. We would like to call the process \textit{virtual photon dissociation}. 



\begin{figure}[h]
	\begin{center}
		\includegraphics[width=8cm]{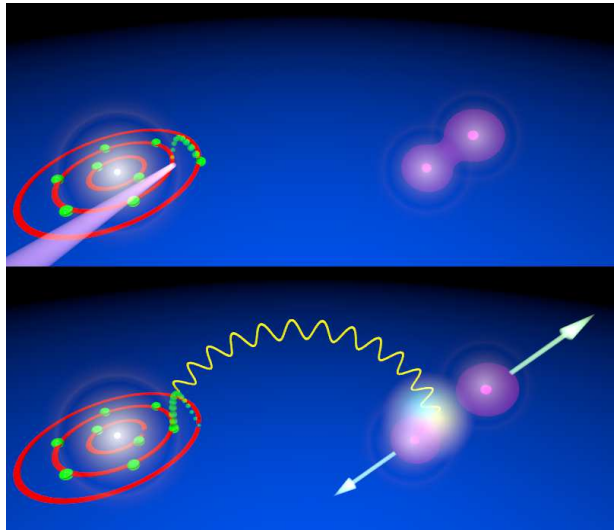}
	\caption{Schematic picture of the dissociation of the molecule after excitation of the neighbor. In the upper panel, the neighbor is electronically excited. The excitation can be by light, but also by other means, like the impact of an electron or ion. If the neighbor is a molecule, the excitation can also be vibrational, see example in the text. The lower panel shows the decay of the neighbor by the emission of a virtual photon which dissociates the molecule. The figure is by courtesy of Till Jahnke.} 
\label{fig:VPD_Excitation}
	\end{center}
\end{figure}


Above, we illuminated the process at hand from the point of view of the neighbor. The molecule undergoes dissociation by the virtual photon emitted from the neighbor and we may ask how large is the corresponding cross section. For this purpose we have to include the excitation process of the neighbor and consider the following scenario: The neighbor, for simplicity an atom, is excited by a photon of momentum $k_{ph}$ and the formed resonance state can decay either by dissociating the molecule or radiatively, the partial widths being $\Gamma_{_{VPD}}$ and $\Gamma_{ph} = \hbar A^{N}$, respectively. The virtual photodissociation cross section can be cast in the form \cite {ICEC_PRA,Taylor_Book}: 
\begin{align}\label{Virtual_CS}
\sigma_{_{VPD}}^M = \frac{\pi}{k_{ph}^2} \frac{g_{d}}{g_{i}} \frac{\Gamma_{_{VPD}} \Gamma_{ph}}{(E_{ph}-E)^{2} + \Gamma^2/4},
\end{align}
where $E_{ph}=\hbar k_{ph} c$ is the energy of the absorbed photon, $E$ is the excess energy introduced above, and $g_{d}$ and $g_{i}$ are the weights of the decaying and initial states. As usual, the total width of the resonance $\Gamma = \Gamma_{_{VPD}} + \Gamma_{ph}$ enters the above Breit-Wigner form. 

To asses the relative size of the virtual and usual cross sections, we make use of the finding that $\Gamma_{_{VPD}}$ in Eq. (\ref{Gamma}) contains the latter. The virtual cross section peaks at $E_{ph}=E$, and at that energy we can conveniently express Eq. (\ref{Virtual_CS}) to give the desired quotient of both cross sections:
\begin{align}\label{Quotient}
\frac {\sigma_{_{VPD}}^M}{\sigma_{_{PD}}^M} = \frac{3}{{(2\pi)}^6} \frac{g_{d}}{g_{i}}\left(\frac{\Gamma_{ph}}{\Gamma_{ph} + \Gamma_{_{VPD}}}\right)^2 \left(\frac{\lambda_{ph}}{R}\right)^6,
\end{align}
where $\lambda_{ph} = 2\pi/k_{ph}$ is the wavelength of the photon. This appealing expression has interesting limits. In particular, if $\Gamma_{ph} \gg \Gamma_{_{VPD}}$, the above quotient is largest and determined solely by the geometric factor $\left(\lambda_{ph}/R\right)^6$. To better understand this, at first sight counterintuitive result, one must keep in mind that the probability for the photon to be absorbed by the neighbor is determined by $\Gamma_{ph}$ and without this absorption the virtual photon process does not take place. 

There are several mechanisms of photodissociation and sizable cross sections are obtained for photon energies reaching excited electronic states of the molecule \cite{Schinke_Book,Rau_Book,Helen_Book,Salman_Book,Bio_Book,PD_Gas,PD_AA}. The cross section for excitation of a vibrational level of the electronic ground state to the dissociation continuum of that state are usually vanishingly small. Can the presence of a neighbor enhance the cross section of the latter case substantially? Let us discuss an explicit example where all the data needed for the calculation is available in the literature. There has been much experimental and theoretical interest in HeH$^{+}$ which is the simplest hetereonuclear two-electron system made of the two most abundant elements in the universe, see \cite{PD_HeH_1,PD_HeH_2} and references therein. The cross section $\sigma_{_{PD}}^M$ for the removal of a proton from the ground state at photon energies below the first excited electronic state has been computed and found to be somewhat smaller than 10$^{-6}$~Mb at the dissociation threshold of 1.844~eV, a tiny quantity indeed \cite{PD_HeH_2}. As a neighbor we choose a Li atom whose 2s $\rightarrow$ 2p excitation is $E=1.85$~eV and has an Einstein coefficient of $A^{N}=3.7\times 10^7$~s$^{-1}$ \cite{NIST}.

Using Eq.~(\ref{Gamma}), one readily obtains $\Gamma_{_{VPD}}=6.05\times 10^{-7}$~cm$^{-1}$ at $R=1$~nm. Since the radiative width  $\Gamma_{ph} = \hbar A^{N} = 1.96\times 10^{-4}$~cm$^{-1}$ is well larger, the quotient of the two cross sections in Eq.~(\ref{Quotient}) is determined by the geometric factor and takes on the very large value ${\sigma_{_{VPD}}^M}/{\sigma_{_{PD}}^M}=4.4\times 10^{12}$. This leads to ${\sigma_{_{VPD}}^M}$ of about 10$^{6}$~Mb !
Even at the large distance $R=100$~\AA{}  between HeH$^+$ and Li, the cross section due to the virtual photon dissociation is still about 1~Mb. Of course, the enormous enhancement only persists in a narrow Breit-Wigner peak around $E_{ph}=1.85$~eV.

The dissociation of systems containing rare gases has been widely studied. Examples are rare gas dimer and trimer ions, rare gas complexes with halogen molecules and with aromatics, see, e.g., \cite{PD_Rg3_Ions,Predissociation_Rg-X2,Rg_Aromatics} and references therein. The available photodissociation investigations essentially relate to excited electronic states because the cross sections at photon energies below those states are usually too small to be measured. The binding of rare gas atoms in neutral systems is typically weak or even very weak. For instance, the binding energy of the NeAr dimer is just 40~cm$^{-1}$ \cite{NeAr_CC,NeAr_ICD}, that of Ar to the aromatic 1-naphtol is 474~cm$^{-1}$ \cite{Rg_Aromatics} and it is even small, 637~cm$^{-1}$ \cite{NeAr_ion} in the NeAr$^+$ ion. How to substantially enhance dissociation in the electronic ground state of these systems? Due to the low binding energy, electronic excitations of the neighbor are not suitable. We can, however, make use of the fact that Eq.~(\ref{Gamma}) and hence also Eq.~(\ref{Quotient}) are not only valid for electronic, but also for vibrational energy transfer \cite{Vibr_ICD}. In other words, we can use these equations to describe the energy transfer from a vibrational level of the neighbor to photoionize the molecule via a virtual photon. If, for example, we take HCN as a neighbor, its bending frequency is 712~cm$^{-1}$ \cite{Radzig_Book} sufficing to dissociate even the NeAr$^+$ ion. Although the radiative width $8\times10^{-11}$~cm$^{-1}$ \cite{Radzig_Book} is small, $\Gamma_{_{VPD}}$ which also contains this term as well as the tiny $\sigma_{_{PD}}^M$, see Eq.~(\ref{Gamma}), can also be small. There is no data on $\sigma_{_{PD}}^M$ available, but assuming that it is similar to the value 10$^{-6}$~Mb of HeH$^{+}$ discussed above, we obtain $\Gamma_{_{VPD}}=4.7\times 10^{-8}$~cm$^{-1}$ at $R=1$~nm. According to (\ref{Quotient}), we now get ${\sigma_{_{VPD}}^M}/{\sigma_{_{PD}}^M}=1.1\times 10^{15}$. An enormous value indeed. 

After having seen that the typically very small dissociation in the electronic ground state can be enhanced dramatically by the presence of a suitable neighbor, we now consider standard molecules at energies where  phtodissociation is substantial. Our examples are the nitrogen (N$_{2}$), water (H$_{2}$O) and methane (CH$_{4}$) molecules which, being common molecules of interest, have been much studied, see, e.g., \cite{PD_N2_1,PD_N2_2,PD_N2_3,PD_H2O_1,PD_H2O_2,PD_H2O_3,PD_CH4_1,PD_CH4_2,PD_CH4_3,PD_AA} and references therein. As neighbors we choose rare gas atoms, which are of interest by themselves, often serve experiments as matrices to trap and investigate molecules \cite{Matrix_Isolation}, and, importantly, the required data to evaluate Eq. (\ref{Gamma}) is available for them.

Photodissociation data on small molecules are beneficially compiled in \cite{PD_AA} and is used here. The photodissciation spectrum of CH$_{4}$ is continuous from below 140~nm photon wavelength to above the ionization potential at 98~nm (12.65~eV). The low excitation energies of Ar, Kr and Xe fall into this range. The wavelength of the 3s$^2$3p$^5$($^2$P$^0_{1/2}$)4s $\rightarrow$ 3s$^2$3p$^6$ transition in Ar is 104.82~nm and the Einstein coefficient is $5.3\times 10^{8}$~s$^{-1}$ \cite{NIST}. At this wavelength $\sigma_{_{PD}}^M = 30$~Mb. With the aid of Eq.~(\ref{Gamma}) we readily find $\Gamma_{_{VPD}} = 0.16$~cm$^{-1}$ at a distance of 1~nm between Ar and the center of mass of methane. This implies that the lifetime of the isolated excited Ar decreases by more than two orders of magnitude from $\tau_{ph} = 1.9$~ns to $\tau_{_{VPD}} = 33$~ps. If we assume that the Ar was excited by a photon, the dissociation via the virtual photon exceeds that of $\sigma_{_{PD}}^M$ by five orders of magnitude. The findings with Kr and Xe as neighbors are similar.  

H$_{2}$O possesses a continuous photodissociation spectrum from below 180~nm to above its ionization threshold at 79.6~nm (15.58~eV) which below 125~nm is accompanied by many peaks. The 4s$^2$4p$^5$($^2$P$^0_{3/2}$)5s $\rightarrow$ 4s$^2$4p$^6$ transition of Kr and 5s$^2$5p$^5$($^2$P$^0_{1/2}$)6s $\rightarrow$ 5s$^2$5p$^6$ of Xe have similar energies, 128.58~and 129.56~nm, and also similar Einstein  coefficients, $3.0\times 10^{8}$~s$^{-1}$ and $2.5\times 10^{8}$~s$^{-1}$, respectively \cite{NIST}. At these wavelengths $\sigma_{_{PD}}^M$ is approximately 8~Mb. With Xe (Kr) as a neighbor at a distance of 1~nm we thus obtain $\Gamma_{_{VPD}} = 0.046$~cm$^{-1}$ (0.05~cm$^{-1}$) which corresponds to a lifetime $\tau_{_{VPD}} = 116$~ps (100~ps) due to the virtual photon dissociation.

For CH$_{4}$ and H$_{2}$O we see that the dissociation by the virtual photon emitted by the neighbor is efficient at the excess energy deposited in the neighbor. In our last example we follow a different scenario. Here, we wish to show that one can avoid having an enhancement of dissociation at a specific incoming photon energy only. A cartoon describing the scenario is shown in Figure 2. To be specific, we consider N$_{2}$ with Ar as a neighbor. The threshold for 3s ionization of Ar is 29.24~eV \cite{Photo_3s_Ar} and thus any photon of wavelength larger than 42.4~nm may produce an Ar$^{*+}$ ion in the state 3s$^1$3p$^6$ which has an excess energy of 13.48~eV \cite{NIST}, well below the ionization potential (15.65~eV) of N$_{2}$. The photodissociation spectrum of N$_{2}$ is dense between about 95 and 85~nm with many intense peaks rising up to $8\times10^{3}$~Mb and at the transition 3s$^1$3p$^6$ $\rightarrow$ 3s$^2$3p$^5$($^2$P$^0_{3/2}$), i.e., at 91.976~nm, the cross section is larger than 150~Mb. The Einstein coefficient for this transition is $1.4\times 10^{8}$~s$^{-1}$ \cite{NIST}.



\begin{figure}[h]
	\begin{center}
		\includegraphics[width=8cm]{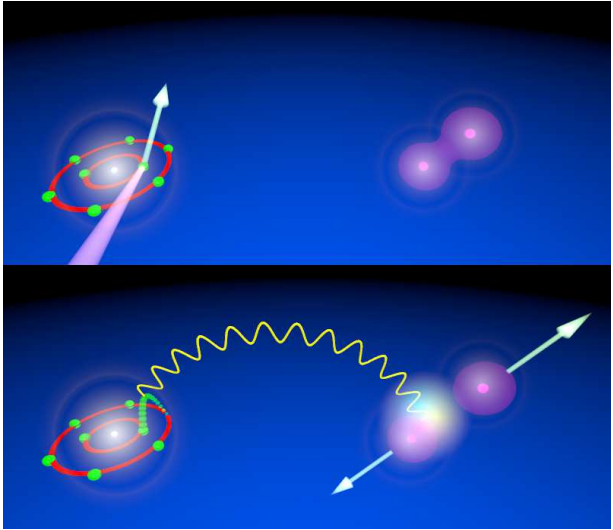}
		\caption{Schematic picture of the dissociation of the molecule after ionization of the neighbor. In the upper panel, the neighbor is ionized. The ionization can be by light, but also by other means, like the impact of an electron or ion. The lower panel shows the decay of the neighbor by the emission of a virtual photon which dissociates the molecule. If the ionization is by light, note that the virtual photon dissociation process continuously takes place as a function of photon energy once the frequency of the light is above the required ionization energy. The figure is by courtesy of Till Jahnke.} 
		\label{fig:VPD_Ionization}
	\end{center}
\end{figure}


 With the above data we can employ Eq. (\ref{Gamma}) and obtain $\Gamma_{_{VPD}} = 0.122$~cm$^{-1}$ at a distance of 1~nm between Ar and the center of mass of N$_{2}$ implying that instead of decaying radiatively in $\tau_{ph} = 7.1$~ns, the excited Ar$^{*+}$ decays much faster in $\tau_{_{VPD}} = 43.5$~ps by emitting a virtual photon which dissociates N$_{2}$. At $R = 5$~\AA{}, which is still a rather large distance where no bonding between N$_{2}$ and Ar occurs, $\tau_{_{VPD}}$ reduces further and becomes sub ps (0.68~ps). 
 
 The fast relaxation of the Ar$^{*+}$ suggests an experiment. Ar-N$_{2}$ is a well studied cluster with equilibrium distance 3.77 \AA{}, see \cite{Ar_N2_1} and references therein. By using modern techniques, one can measure in coincidence the momenta of the charged particles, see, e.g., \cite{Ne2_Exp,HeNe_Exp}. In the present case, this implies that one can measure the photoelectrons identifying the creation of Ar$^{*+}$ with a hole in 3s in coincidence with the momentum of the relaxed Ar$^+$ in its ground state 3s$^2$3p$^5$. The dissociation of N$_{2}$ should be reflected in this momentum distribution and can be well distinguished from the relaxation of Ar$^{*+}$-N$_{2}$ by photon emission forming Ar$^+$-N$_{2}$. The possibility of measuring in coincidence photons, ions and electrons in decay processes has been demonstrated \cite{Photon_ICD_1,Photon_ICD_2}. So, at least in principle, one could also measure additionally in coincidence the possible photons emitted from the atomic nitrogen fragments formed by the virtual photodissociation. For completeness, we just mention that there are several methods of N-atom product detection by fluorescence and other means, see \cite{N_Detection_1,N_Detection_2,N_Detection_3} and references therein.
  
 \section{Discussion}
 
 We have seen that fragmentation of molecules by virtual photons emitted from remote neighbors can be efficient and fast. The involved energy transferred can be electronic, but it can also be vibrational. Two scenarios have been discussed: excitation or ionization of the neighbor. If the excess energy is deposited by a photon, the enhancement of the photodissociation cross section can be dramatic at the exciting photon energy. In the case of ionization, the dependence on the ionizing photon energy is continuous. The exciting/ionizing particle does not have to be a photon, it can be an electron or an ion. 
 
 We notice that there is an analogy to another process. Interatomic Coulombic decay (ICD) is an efficient decay channel in excited/ionized systems, such as van der Waals and hydrogen bonded clusters and solutions.  In the ICD process the de-excitation of a excited/ionized atom or molecule via energy transfer to the environment causes ionization of the environment through long range electronic correlation.  Since its prediction \cite{PhysRevLett.79.4778}, ICD has been widely studied (see, e.g., \cite{0953-4075-48-8-082001} and references therein), found to be ultrafast (typically on the femtosecond timescale), and in most cases to be fast enough to quench concurrent electronic and nuclear mechanisms \cite{Janke:2010,ICD_H2O_Uwe,:/content/aip/journal/jcp/137/3/10.1063/1.4731636,:/content/aip/journal/jcp/144/16/10.1063/1.4947238}. Although ICD can be purely electronic, e.g., be operative between atoms, and there is no need for nuclear motion, one can learn much from ICD on virtual photon dissociation discussed here. In ICD, retardation tends to enhance the decay rate \cite{Retardation_ICD_1,Retardation_ICD_2} and we may assume that this is also the case here. Excitation of the neighbor by a photon may influence considerably the photoionization cross section of another atom \cite{Two_Center_Carsten} and we have seen that this also applies here for the photodissociation of the molecule. ICD becomes particularly fast when the intermolecular (interatomic) distances are small and the number of involved species is large \cite{ICD_more_N_2,Stumpf16a,:/content/aip/journal/jcp/145/10/10.1063/1.4962353,res_Robin}. Similarly, if several molecules which can be dissociated at the wavelength at hand are available, the excited/ionized neighbor decays faster as the number of decay channels grows. If there are more neighbors, the probability for the impinging particle to excite/ionize one of them, trivially grows and hence also that to dissociate the molecule. We conclude by stressing that virtual photon dissociation and ICD complement each other. If the excess energy suffices to ionize the environment, ICD can take place, and if not, virtual photon dissociation can be operative.

\bibliographystyle{naturemag}
\bibliography{biblio}

\section{Acknowledgements}

\begin{acknowledgements}
	The author thanks K. Gokhberg, T. Jahnke, A. I. Kuleff and A. Saenz for valuable contributions. Financial support by the European Research Council (ERC) (Advanced Investigator Grant No. 692657) is gratefully acknowledged
\end{acknowledgements}

{\bf Competing interests:} The author declares no competing interests.

\newpage

\end{document}